\documentclass{article}

\usepackage{exscale}
\usepackage{latexsym}
\usepackage{amsmath}
\usepackage{amssymb}
\usepackage[OT1]{fontenc}
\usepackage[latin1]{inputenc}
\usepackage{graphics}
\usepackage{graphicx}
\usepackage{color}

\begin{document}
\sloppy

\markboth{Lichtenegger, Schappacher}
{Lichtenegger, Schappacher}

\title{Seeing about Soil -- Management Lessons from a
Simple Model for Renewable Resources}

\date{December 2, 2013}

\author{KLAUS LICHTENEGGER,\\
\small Bioenergy 2020+ GmbH\\
\small Gewerbepark Haag 3, 3250 Wieselburg-Land, Austria\\
\small k\_lichtenegger@yahoo.de\\[12pt]
WILHELM SCHAPPACHER,\\
\small Institute of Mathematics and Scientific Computing, Heinrichstra\ss{}e 36 \\
\small Graz University, 8010 Graz, Austria\\
\small wilhelm.schappacher@uni-graz.at}

\maketitle

\begin{abstract}
\noindent Employing an effective cellular automata model, we investigate and
  analyze the build-up and erosion of soil. Depending on the strategy employed
  for handling agricultural production, in many cases we find a critical dependence
  on the prescribed production target, with a sharp transition between stable
  production and complete breakdown of the system.
  
  Strategies which are particularly well-suited for mimicking real-world
  management approaches can produce almost cyclic behaviour, which
  can also either lead to sustainable production or to breakdown.
  
  While designed to describe the dynamics of soil evolution, this model is
  quite general and may also be useful as a model for other renewable
  resources and may even be employed in other disciplines like psychology.\\[6pt]
\end{abstract}

\section{Introduction}

At present, a large part of our economy is built on the exploitation of nonrenewable resources as, for instance,
oil, coal and natural gas. It is common sense that we have to put more emphasis on the use of renewable resources,
as the amount of nonrenewable resources is strictly limited~\cite{LimitsGrowth, LimitsUpdate}. Even before these
limits are reached, the technological and economic effort of exploitation rises dramatically and ecological
consequences tend to become unacceptably severe.

Thus a fundamental change in our behaviour will be necessary in order to avoid catastrophic breakdowns as well
as ecological disasters. But also resources which are renewable can still be ruined by overexploitation, with
well-known examples for this being overfishing and deforestation.

\medskip

A particularly important resource both for natural systems and for agriculture is soil.
The loss of soil is a topic intensively discussed in recent years, and loss of soil has been made responsible for the
breakdown of whole civilizations~\cite{Dirt}. Apart from the obvious importance for plant growth,
soil is an important ecosystem by itself and one of the largest reservoirs of biodiversity.
Unbalanced soil ecology is assumed to be a key factor for the spread of several diseases~\cite{NYTBiodiv}.
In addition, soil can contain large amounts of carbon and building up soil can be a very efficient strategy
to capture carbon and remove it from the atmosphere, thus slowing down climate change.

Soil quality, build-up and loss are a topic of intense research, but the mathematical models employed
so far tend to be either very simple (for sake of being analytically tractable) or already very specific,
including complex chemical and biochemical reactions and
connections~\cite{CritLevelSOM, CN-SIM1, CN-SIM2}.

In this article we propose an alternative model which captures the main elements of soil evolution.
This model exhibits complex behaviour (including critical transitions) but is, at the same time, simple
enough to be easily applied also to other renewable resources. For these reasons it can serve as a testing
ground for various management strategies.

\section{The Model}
\label{sec:model}

The present model had initially been built to supplement
a climate model presented in~\cite{LiSchapp:2011CarbonClimate}
with a qualitative description of the interplay of soil and plant life,
in particular regarding carbon storage capacity.

Soon it turned out that already without being coupled to
the climate model, the plant-soil system revealed dynamics
well worth of separate investigation, and it became the
topic of the present article.

\subsection{Basic Concept}

The main ideas behind the model are:
\begin{itemize}
  \item Natural vegetation builds up soil.
  \item Natural vegetation is replaced by agricultural fields,
    depending on a prescribed production goal and the chosen
    management strategy.
  \item In agricultural fields, soil is lost due to intense agriculture,
    and below a certain threshold of soil, agricultural production
    starts to decline.
\end{itemize}

\noindent In a rather simplified approach to natural succession, we
introduce two kinds of plant population:

\begin{itemize}
  \item Pioneer plants build up soil slowly, but they are able
    to inhabit even ``empty'' cells (i.e. those without soil)
    and spread randomly.
  \item Forests build up soil faster, but they require a certain
    soil threshold $\tau_{\rm for}$ to grow at all. In addition, forests can
    (except in the case of deliberate reforestation, as discussed in
    sec.~\ref{sssec:resource_management}) only grow if at least one
    neighbor cell is also inhabited by forests.
\end{itemize}

\noindent Different strategies for creating and abandoning agricultural
fields can be employed in order to fulfill the prescribed
production goal, see sec.~\ref{sssec:resource_management}.

\subsection{Implementation}

The model is implemented by means of extended stochastic cellular
automata~\cite{vNThSelfReprod, Lichtenegger:2005DA}. We define the
system on a grid, with states conveniently represented by two matrices:

\begin{itemize}
  \item The inhabitance state of the cells is described
    by the matrix $\boldsymbol{P}(n_t)$ with entries
    \mbox{$p_{ij}(n_t)\in\{0,\,1,\,2,\,3\}$}, where we use
    the following assignment:
    \begin{enumerate}
      \item[$0\ldots$] Cell is uninhabited.
      \item[$1\ldots$] Cell inhabited by pioneer plants
      \item[$2\ldots$] Cell inhabited by forest
      \item[$3\ldots$] Cell used for agriculture
    \end{enumerate}
  \item The amount of soil contained in the cells is
    described by the matrix $\boldsymbol{S}(n_t)$ with entries
    $s_{ij}(n_t)\in\mathbb{R}_0^+$.
\end{itemize}

The parameters of the model are summarized in table~\ref{tab:symbols}.

\begin{table}
\begin{tabular}{|c|c|l|} \hline
symbol & range & meaning/definition \\ \hline
$N_1$ & $\in\mathbb{N}$ & size of grid (number of cells) in first direction\\[3pt]
$N_2$ & $\in\mathbb{N}$ & size of grid (number of cells) in second direction \\[3pt]
$N_t$ & $\in\mathbb{N}$ & total number of timesteps \\[3pt]
$\rho_{\text{for}}^{\text{(ini)}}$ & $\in[0,1]$ & initial density of forests (before placing fields) \\[6pt]
$\rho_{\text{pio}}^{\text{(ini)}}$ & $\in[0,1]$ & initial density of pioneer plants \\[6pt]
$A^{(\rm goal)}$ & $\in [0,\,N_1N_2]$ & goal for agricultural production \\[3pt]
$\gamma_{\rm agr}$ & $\in[0,1]$ & normalized production goal,
  $\gamma_{\rm agr} = \frac{A^{(\rm goal)}}{N_1N_2}$ \\[6pt]
$p_{\text{pio}}$ & $\in[0,1]$ & probability of an empty cell to become populated by pioneer plants \\[6pt]
$\tau_{\text{for}}$ & $\in\mathbb{R}^+$ & soil threshold for forest growth \\[3pt]
$\tau_{\rm agri}$ & $\in\mathbb{R}^+$ & soil threshold for decline of agricultural production \\[3pt]
$s^{(+)}_{(\text{pio})}$ & $\in\mathbb{R}^+$ & soil build-up per timestep by pioneer plants \\[6pt]
$s^{(+)}_{(\text{for})}$ & $\in\mathbb{R}^+$ & soil build-up per timestep by forest \\[6pt]
$s^{(-)}_{(\text{agr})}$ & $\in\mathbb{R}^+$ & soil loss per timestep due to agriculture \\[3pt]
\hline \end{tabular}
\caption{List of parameters employed in the model}
\label{tab:symbols}
\end{table}

\subsubsection{Initialization}

The grid initalization is done in four consecutive steps, which set up the
matrices $\boldsymbol{P}(0)$ and $\boldsymbol{S}(0)$:
\begin{enumerate}
\item Forests, described by $p_{ij}(0)=2$, are distributed in the matrix
  $\boldsymbol{P}(0)$ with the initial forest density $\rho_{\text{for}}^{\text{(ini)}}$.
\item Soil values are determined according to
   \begin{equation}
     s_{ij}(0) = \tau_{\text{for}}\,\left( \delta_{2,p_{ij}(0)} + u_{ij} \right)
   \end{equation}
   with $\delta_{k\ell}$ denoting the Kronecker delta
   and $u_{ij}$ taken randomly from a uniform $[0,\,1]$-distribution.
\item Agricultural fields are placed employing one of the
   strategies described in sec.~\ref{sssec:resource_management},
   possibly making use of the information contained in $\boldsymbol{S}(0)$.
   Note that this will typically reduce the density of forest below
   the initial value $\rho_{\text{for}}^{\text{(ini)}}$.
\item Of the remaining cells with $p_{i,j}(0)=0$, a fraction of
   $\rho_{\text{pio}}^{\text{(ini)}}$ is populated with pioneer plants,
   $p_{i,j}(0)\to 1$.
\end{enumerate}
The target number of agricultural cells is initialized as
$N_{\text{agr}}=A^{(\rm goal)}$.

\subsubsection{Update Rules}

At each time update $n_{t}\to n_{t}+1$ the following update rules are executed:
\begin{itemize}
\item According to the re-population probability $p_{\text{pio}}$, empty cells
  are populated with pioneer plants.
\item Any cell $(i,j)$ which is either empty or populated with pioneers is
  converted to forest with probability
  \begin{equation}
    p = \frac{\text{number of neighbouring forest cells}}{\text{total number of neighbour cells}}
    \label{eq:update_forest}
  \end{equation}
  if the condition $s_{ij}(n_t)\ge \tau_{\text{for}}$ is met. Neighbourhood is
  defined as a mixed von Neumann-Moore type, with diagonal neighbours
  carrying half the weight of directly adjacent ones. (In terms of~\cite{LiSchapp:2009Forest},
  this is implemented by setting $\pi_{\rm M}=\frac12$.)
\item Agricultural fields are created or abandoned according to the current
  management strategy, as described in sec.~\ref{sssec:resource_management}.
\item The agricultural production $A(n_t)$ is
  calculated as decribed in~\ref{sssec:agricultural_production} and the new target
  number $N_{\text{agr}}$ is determined according to chosen strategy, based
  on the relation between production $A(n_t)$ and production goal $A^{(\rm goal)}$.
\item Soil content is modified according to the current cell population,
  \begin{align}
    \delta s(n_t) & = s^{(+)}_{(\text{pio})}\delta_{1,p_{ij}(n_t)}
      + s^{(+)}_{(\text{for})}\delta_{2,p_{ij}(n_t)}
      - s^{(-)}_{(\text{agr})}\delta_{3,p_{ij}(n_t)},\\
    s_{ij}(n_t+1) & = \max\left(0, s_{ij}(n_t) + \delta s(n_t) \right),
  \end{align}
  which model soil generation and erosion.
\item Cells with $s_{ij}(n_t)=0$ become empty, $p_{ij}(n_t+1)=0$.
\end{itemize}

\subsubsection{Determining Agricultural Production}
\label{sssec:agricultural_production}

The agricultural production $a$ of an appropriate cell $(i,\,j)$ during
timestep $n_t$ is equal to a constant $a_0$ as long as the amount of soil is
above a fixed threshold, $s_{ij}(n_t) \ge \tau_{\rm agri}$.
If $s_{ij}(n_t) < \tau_{\rm agri}$, we assume direct proportionality of production
to the amount of soil, with a continuous transition between the two regimes.
The constant $a_0$ can be chosen equal to one, which normalizes the
production and thus sets the scale. This yields
\begin{equation}
  a(i,j,n_t) =  \frac{\delta_{3,p_{i,j}(n_t)}}{\tau_{\rm agri}}\,\min\left( s_{ij}(n_t),\,\tau_{\rm agri}\right).
\end{equation}
The total production is obtained as
\begin{equation}
  A(n_t) = A(\boldsymbol{P},\boldsymbol{S},n_t) = \sum_{i,j} a(i,j,n_t).
\end{equation}

\subsubsection{Resource Management Strategies}
\label{sssec:resource_management}

The most intricate component of the model is the strategy how to handle
the agricultural fields. A multitude of strategies is possible, of which we
have chosen a small subset which is relatively easy to describe and to
implement.

All strategies studied in this article are based on four basic ways to
handle the distribution of agricultural cells, which for brevity are denoted
as "fields":
\begin{enumerate}
  \item[{[S1]}] Keep fields as long as possible
  \item[{[S2]}] Abandon fields as soon as soil drops below agricultural threshold
  \item[{[S3]}] Redistribute fields at each step, no reforestation
  \item[{[S4]}] Redistribute fields at each step, reforestation if possible
\end{enumerate}
Reforestation is the only way how a cell can be populated with forest without
having a neighbour forest cell. Reforestation of a cell $(i,j)$ only succeeds if
$s_{ij}(n_t)\ge \tau_{\rm for}$.
These basic strategies have been combined with two
ways how cells are chosen as new fields:
\begin{itemize}
  \item Choose new fields randomly: ${}_{\text{rand}}$,
  \item Choose new fields according to maximum soil: ${}_{\text{opt}}$
\end{itemize}
In addition we compare two ways how to handle
the number of agricultural fields:
\begin{itemize}
  \item Number of fields is held fixed: ${}^{\text{(fix)}}$,
  \item Number of fields is variable: ${}^{\text{(var)}}$
\end{itemize}

In the present simulation, for ${}^{\text{(var)}}$ we have employed the strategy
to raise the number of fields by $1\%$,
if during the last timestep the production was below the production
goal, $A(n_t-1)<A^{(\rm goal)}$ and abandon $1\%$ of the fields if
$A(n_t-1)>A^{(\rm goal)}$. The rate of $1\%$ per timestep is
arbitrary, but it at least qualitiatively captures the inertia effects
induced by the effort of establishing new fields and by the reluctance
of abandoning existing ones.

\medskip

\noindent Combinatorics yields sixteen different strategies how to manage agricultural
production within this model. For example strategy $\text{[S1]}_{\text{rand}}^{\text{(fix)}}$
would keep fields as long as possible, but keep the number of fields fixed. With
this strategy certain drops in productivity are unavoidable within this model.

\smallskip

In sec.~\ref{sec:comparison} it will turn out that depending on the production goal,
the same principal strategy can yield dramatically different results.

\section{Sample Runs and Methods of Data Analysis}
\label{sec:sample_analysis}

\begin{figure}
  \includegraphics[width=12cm]{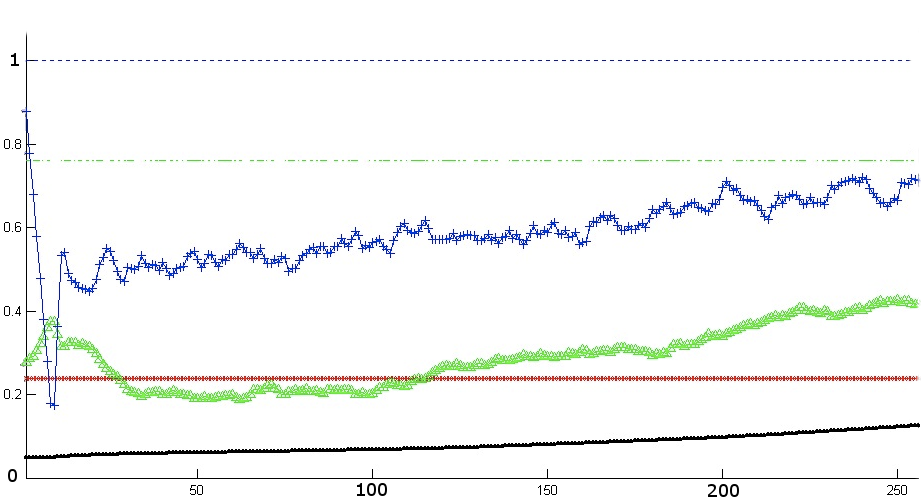}\put(-330,168){(a)}\put(-1,5.5){$\rightarrow$}\put(0,0){$n_t$}\\
  \includegraphics[width=12cm]{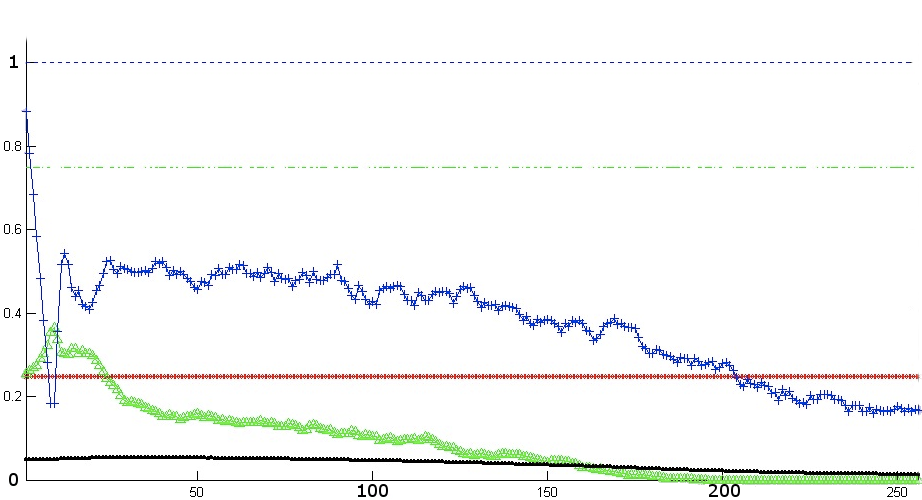}\put(-330,168){(b)}\put(-1,5.5){$\rightarrow$}\put(0,0){$n_t$}
  \caption[Influence of production target on simulation results, illustrated by sample simulation
    runs.]{Influence of production target on simulation results, illustrated by
    sample simulation runs. For these runs, strategy $\text{[S1]}_{\text{rand}}^{\text{(fix)}}$
    has been employed with (a) $\gamma_{\rm agr}=0.24$ and (b) $\gamma_{\rm agr}=0.25$.\\
    Other parameters have been set to standard values as given in tab.~\ref{tab:standardpar}.\\
    We plot the following quantities as function of the time step index $n_t$:\\[3pt]
    \begin{tabular}{cl}
    \raisebox{-1mm}{\includegraphics[height=3mm]{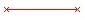}} & density of agricultural cells,\\
    \raisebox{-1mm}{\includegraphics[height=3mm]{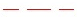}} & minimum possible density of agricultural cells,\\
    \raisebox{-1mm}{\includegraphics[height=3mm]{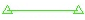}} & density of forest cells,\\
    \raisebox{-1mm}{\includegraphics[height=3mm]{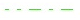}} & maximum possible density of forest cells,\\
    \raisebox{-1mm}{\includegraphics[height=3mm]{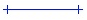}} & normalized agricultural production,\\
    \raisebox{-1mm}{\includegraphics[height=3mm]{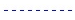}} & normalized target agricultural production (equal to one),\\
    \raisebox{-1mm}{\includegraphics[height=3mm]{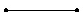}} & amount of soil, displayed on a linear-logarithmic scale (see eq.~\eqref{eq:soil_linlogscale}).
    \end{tabular}}
  \label{fig:run32-24-25}
\end{figure}

A typical sample run for the model, employing strategy
$\text{[S1]}_{\text{rand}}^{\text{(fix)}}$ and standard parameters
as given in table~\ref{tab:standardpar}, is shown in fig.~\ref{fig:run32-24-25}.a.

\medskip

\noindent The plot shows the density of agricultural cells
\begin{equation}
  \rho_{\text{agr}}(n_t) = \frac1{N_1N_2}\sum_{i,j}\delta_{3,p_{ij}(n_t)},
\end{equation}
(which is constant in this case), the density of trees
\begin{equation}
  \rho_{\text{for}} = \frac1{N_1N_2}\sum_{i,j}\delta_{2,p_{ij}(n_t)},
\end{equation}
the normalized agricultural production
\begin{equation}
  a(n_t) = \frac{A(n_t)}{A^{(\rm goal)}}
\end{equation}
and a measure for the amount of soil. Since the average amount of soil can grow
very large in this model, we do not directly plot the normalized amount of soil
\begin{equation}
  s(n_t)=\frac1{N_1N_2}\sum_{i,j}s_{ij}(n_t),
\end{equation}
but instead show the quantity
\begin{equation}
  \tilde s(n_t) = \ln\left( 1 + \frac{s(n_t)}{s_0} \right)
  \label{eq:soil_linlogscale}
\end{equation}
with the (arbitrary) reference scale $s_0=10$.
This yields an approximately linear relation $\tilde s \sim \frac{s}{s_0}$ for $s\ll s_0$
and an approximately logarithmic relation $\tilde s\sim \ln\frac{s}{s_0}$ for $s\gg s_0$.

\begin{table}
  \begin{center}
  \begin{tabular}{|llll|} \hline
  type of parameter & \multicolumn{3}{c|}{parameters and values} \\[3pt] \hline
  grid size \& timesteps & $N_1=32$, & $N_2=32$, & $N_t=256$, \\[3pt]
  initial conditions: &
  $\rho_{\text{for}}^{\text{(ini)}} = 0.5$, &
  $\rho_{\text{pio}}^{\text{(ini)}} = 0.5$, & \\[6pt]
  probabilities \& thresholds: &
  $p_{\text{pio}} = 0.5$, &
  $\tau_{\text{for}} = 0.5$, &
  $\tau_{\rm agri} = 1$, \\[6pt]
  soil addition/subtraction: &
  $s^{(+)}_{(\text{pio})} = 0.025$, &
  $s^{(+)}_{(\text{for})} = 0.05$, &
  $s^{(-)}_{(\text{agr})} = 0.1$ \\ \hline
\end{tabular}
\end{center}
\caption{Standard parameters used for most simulation in
secs.~\ref{sec:sample_analysis} and~\ref{sec:comparison}}
\label{tab:standardpar}
\end{table}

As expected for strategy $\text{[S1]}_{\text{rand}}^{\text{(fix)}}$, in
fig.~\ref{fig:run32-24-25} within the first few timesteps there is steep
drop in agricultural production. While the density of trees increases
during these steps, it decreases when agricultural fields get
abandoned and new ones are created.

In this simulation run, both the agricultural production and the density
of forests almost stabilize after about 30 timesteps and start to slowly
increase after about 100 steps.

\bigskip

\noindent This picture dramatically changes when raising the agricultural
production goal from $\gamma_{\rm agr}=0.24$ to $\gamma_{\rm agr}=0.25$.
A typical simulation run for this goal for otherwise unchanged parameters is
shown in fig.~\ref{fig:run32-24-25}.b.

While the time evolution is quite similar during the first 30 time steps,
it begins to take a different path afterwards. The density of forests
does not stabilize on a finite level but slowly drops to zero. The
agricultural production does not recover but -- except for fluctuations --
slowly drops and may only stabilize on a rather low level after loss of
all forest cells.

\subsection{Long-Term Behaviour of the Model}
\label{ssec:longterm}

\begin{figure}
  \includegraphics[width=12cm]{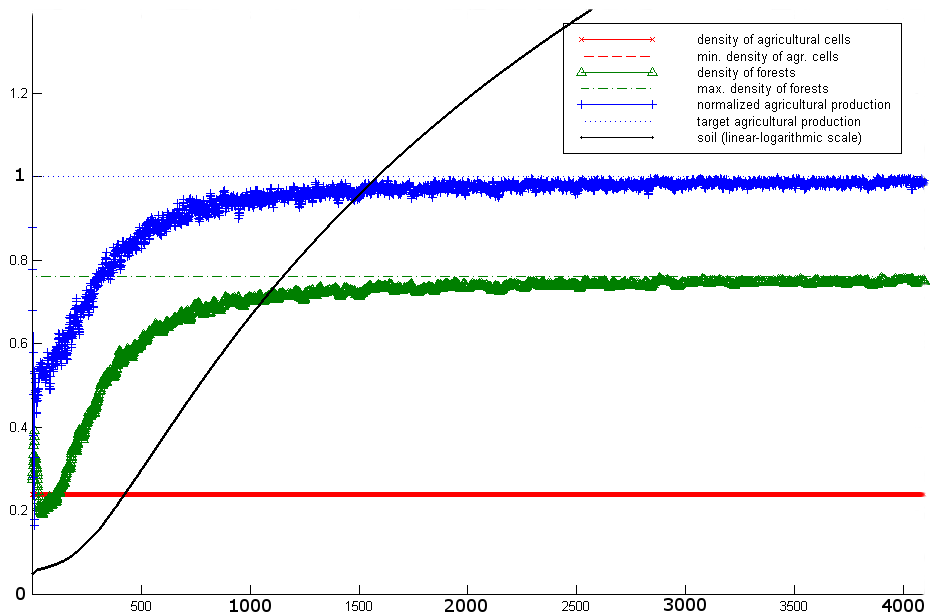}\put(-326,210){(a)}\put(-6,5){$\rightarrow$}\put(-4,-2){$n_t$}\\
  \includegraphics[width=12cm]{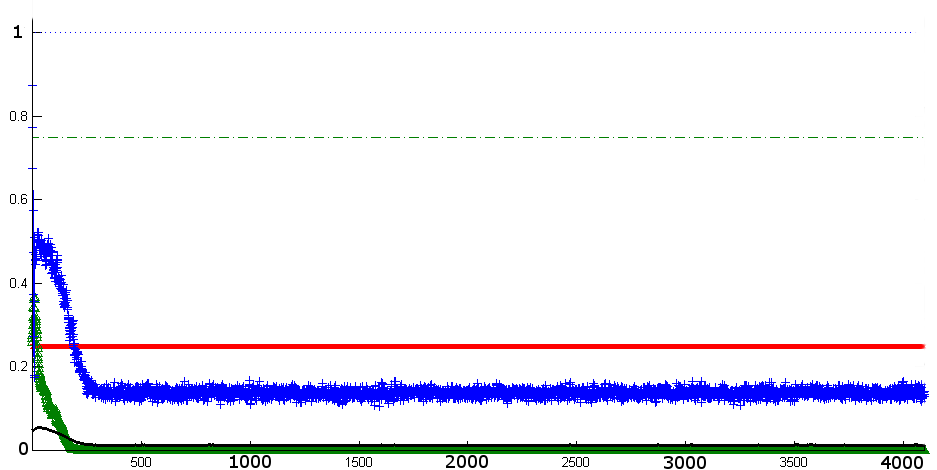}\put(-326,168){(b)}\put(-6,4.6){$\rightarrow$}\put(-4,-2){$n_t$}
  \caption[Illustration of long-term behaviour]{Illustration of long-term behaviour:\\
  Sample simulation runs as in fig.~\ref{fig:run32-24-25}
  (strategy $\text{[S1]}_{\text{rand}}^{\text{(fix)}}$, parameters from
  tab.~\ref{tab:standardpar}), except of the total simulation length $N_t=4096$. 
  Run performed with (a) $\gamma_{\rm agr}=0.24$, (b) $\gamma_{\rm agr}=0.25$.
  See fig.~\ref{fig:run32-24-25} for a more detailed explanation
  of the quantities displayed in the plot.}
  \label{fig:longterm}
\end{figure}
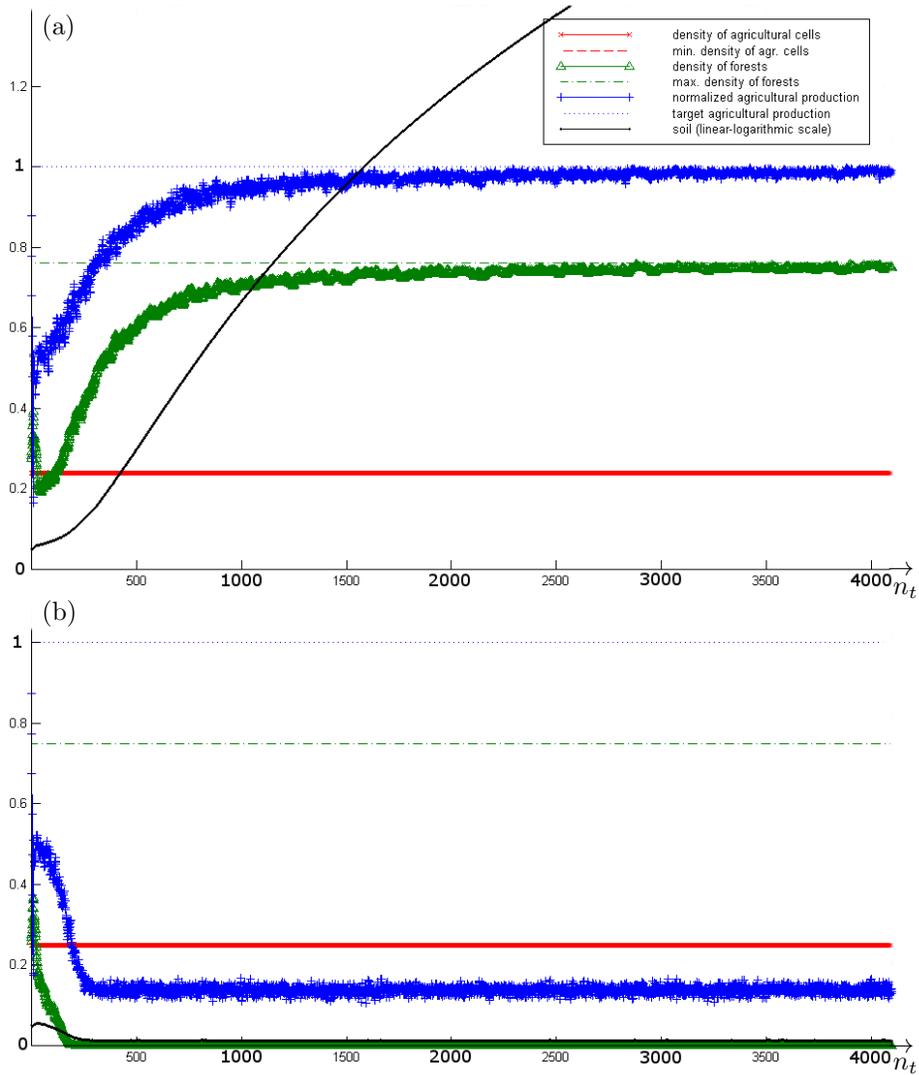

While within $N_t=256$ timesteps the trends already become quite clear,
it is interesting to check the long-term behaviour as well. Typical runs for
the same choice of parameters as in fig.~\ref{fig:run32-24-25} but for
$N_t=4096$ timesteps are shown in fig.~\ref{fig:longterm}.

Performing a larger number of runs for different sets of parameters
shows that these are the two main outcomes of model runs: Either a
positive level of forests is maintained and the agricultural
production stabilizes at or close to the target production or forests vanish
due to over-exploitation and the agricultural production breaks down to
stabilize at a very low level, in particular significantly below the target
production level.

\subsection{Estimating Finite-Size Effects}
\label{ssec:finite_size}

\begin{figure}
  \includegraphics[width=12cm]{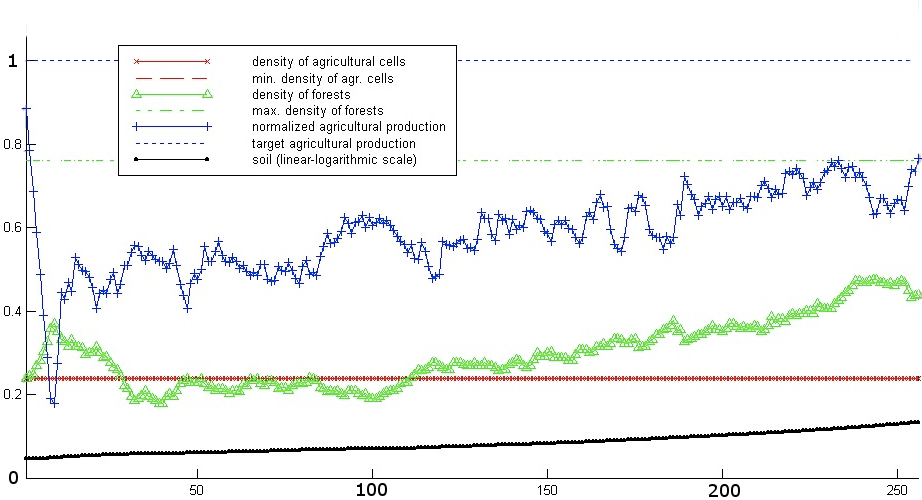}\put(-326,168){(a)}\put(-5,5.5){$\rightarrow$}\put(-3,-2){$n_t$}\\
  \includegraphics[width=12cm]{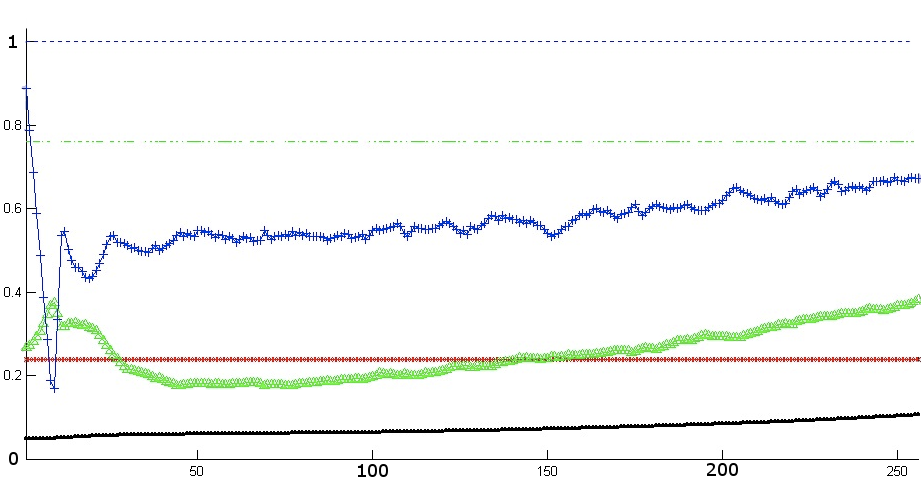}\put(-326,168){(b)}\put(-5,5.2){$\rightarrow$}\put(-3,-2){$n_t$}
  \caption[Estimate of finite-size effects]{Estimate of finite-size effects:
  Sample simulation runs as in fig.~\ref{fig:run32-24-25}
  (strategy $\text{[S1]}_{\text{rand}}^{\text{(fix)}}$, parameters from tab.~\ref{tab:standardpar}), 
  with $\gamma_{\rm agr}=0.24$, except\\
  (a) $N_1=N_2=16$, (b) $N_1=N_2=64$.\\
  See fig.~\ref{fig:run32-24-25} for a more detailed explanation of the quantities displayed in the plot.}
  \label{fig:run1664-24}
\end{figure}
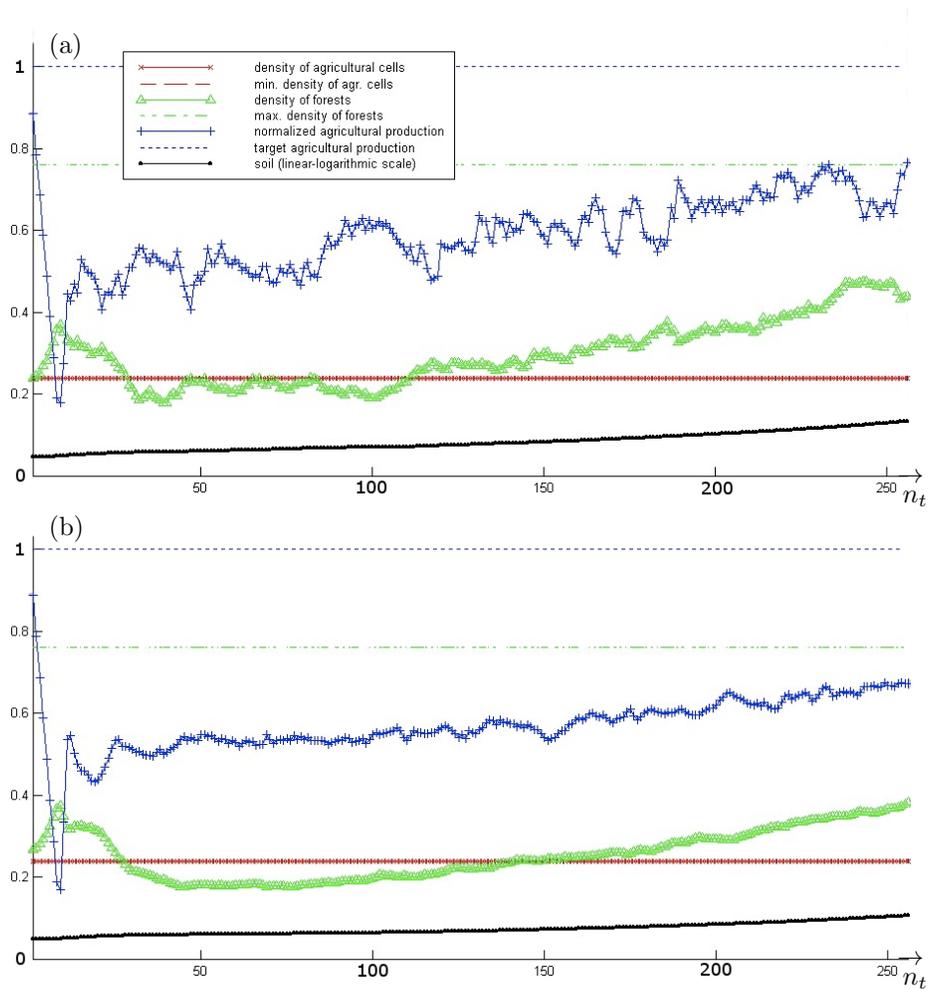

The simulations shown in figs~\ref{fig:run32-24-25} to~\ref{fig:longterm}
have been performed on a $[32\times32]$-grid. One can expect such
simulations to be significantly affected by finite-size effects. In order
to have at least a qualitative estimate for the severity of these effects,
in fig.~\ref{fig:run1664-24} we show the results for two simulations with
parameters chosen as in fig.~\ref{fig:run32-24-25}.a except that one has been
performed on a $[16\times16]$-, the other one on a $[64\times64]$-grid.

While for smaller grids the fluctuations (mostly induced by the stochastic
nature of the model) are more pronounced, the general behaviour of the
model seems to be quite insensitive to the size of the grid.

\subsection{Dependence on Target Production Rate}
\label{ssec:dep_prodrate}

While single simulation runs as depicted in figs.~\ref{fig:run32-24-25}
to~\ref{fig:run1664-24} give some hints on the characteristics of
the model, a more systematic analysis is required for deeper insight.
Therefore we have defined production target values
\[
  \gamma_{\rm agr}\in0.01\cdot\{1,\,2,\,\ldots,\,34\}
\]
and for each value performed 20 simulations with strategy
$\text{[S1]}_{\text{rand}}^{\text{(fix)}}$ and parameters
given in tab.~\ref{tab:standardpar}.

The maximum value $\gamma_{\rm agr}=0.34$ has been chosen
because the maximum sustainable normalized agricultural
production $\gamma_{\rm max}$, determined by the linear equation
\begin{equation}
  (1-\gamma_{\rm max})\,\max\left\{s^{(+)}_{(\text{pio})},\,s^{(+)}_{(\text{for})}\right\}
  - \gamma_{\rm max}\,s^{(-)}_{(\text{agr})} = 1,
  \label{eq:determine_gammamax}
\end{equation}
is equal to $\frac13$ for the parameter values employed in the simulation.

In order to obtain a small number of characteristic numbers for each simulation
run, we have recorded the density of agricultural cells, the density of forests, the
normalized agricultural production and the amount of soil for the last
$\frac{N_t}4=64$ timesteps, assuming that after $\frac{3N_t}4=192$ steps
in most cases a characteristic state has been reached.

\begin{figure}
  \includegraphics[width=12cm]{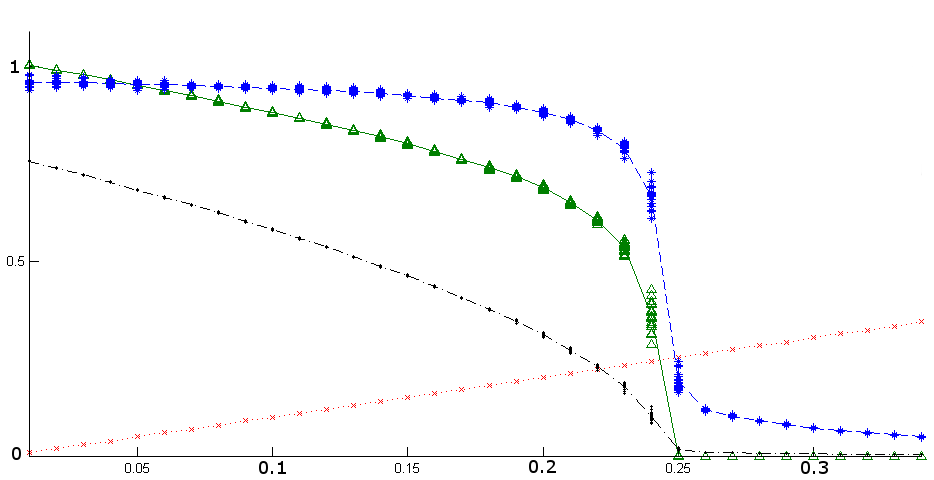}\put(-6,6){$\rightarrow$}\put(-20,0){$\gamma_{\rm agr}$}
  \caption[Structure of an overview plot]{Typical structure of an overview plot:
     Summary of simulation runs for strategy $\text{[S1]}_{\text{rand}}^{\text{(fix)}}$
     and standard parameters given in tab.~\ref{tab:standardpar}.
     For each value of $\gamma_{\rm agr}\in 0.01\cdot\{1,\,2,\,\ldots,\,34\}$,
     20 single simulations have been performed and the the values of characteristic quantities,
     averaged over the last 64 timesteps have been recorded. We show both the average for all
     20 runs and the single results (in order to give an impression of the characteristic deviations).
      We plot the following quantities as function of the the target production rate $\gamma_{\rm agr}$:\\[3pt]
    \begin{tabular}{cl}
    \raisebox{-1mm}{\includegraphics[height=3mm]{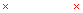}} & density of agricultural cells for the single runs,\\
    \raisebox{-1mm}{\includegraphics[height=3mm]{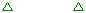}} & density of forest cells for the single runs,\\
    \raisebox{-1mm}{\includegraphics[height=3mm]{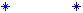}} & normalized agricultural production for the single runs,\\
    \raisebox{-1mm}{\includegraphics[height=3mm]{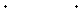}} & soil content (lin.-log. plot) for the single runs, \\
    \raisebox{-1mm}{\includegraphics[height=3mm]{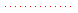}} & density of agricultural cells, average value of all runs,\\
    \raisebox{-1mm}{\includegraphics[height=3mm]{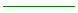}} & density of forest cells, average value of all runs,\\
    \raisebox{-1mm}{\includegraphics[height=3mm]{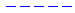}} & normalized agricultural production, average value of all runs,\\
    \raisebox{-1mm}{\includegraphics[height=3mm]{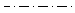}} & soil content (lin.-log. plot), average value of all runs
    \end{tabular}}
  \label{fig:strat1rf}
\end{figure}

The plot of results is shown in fig.~\ref{fig:strat1rf}. The dramatic transition
between $\gamma_{\rm agr}=0.24$ and $\gamma_{\rm agr}=0.25$, already
to be expected from a comparison of figs.~\ref{fig:run32-24-25}.a and~\ref{fig:run32-24-25}.b,
is clearly visible here. The breakdown of agricultural production is accompanied
by a steep drop of the density of forests. Only in the region $0.23\le\gamma_{\rm agr}\le 0.25$
there is a significant spread of the single results, which will is examined in more detail and
discussed in sec.~\ref{ssec:analysis_transition}.

\subsection{Analysis of the Transition Region}
\label{ssec:analysis_transition}

The overview plot shown in fig.~\ref{fig:strat1rf} and discussed in sec.~\ref{ssec:dep_prodrate}
clearly shows that for strategy $\text{[S1]}_{\text{rand}}^{\text{(fix)}}$ one has a significant
spread of results only in a narrow transition region $0.23\le\gamma_{\rm agr}\le 0.25$.

\begin{figure}
  \includegraphics[width=12cm]{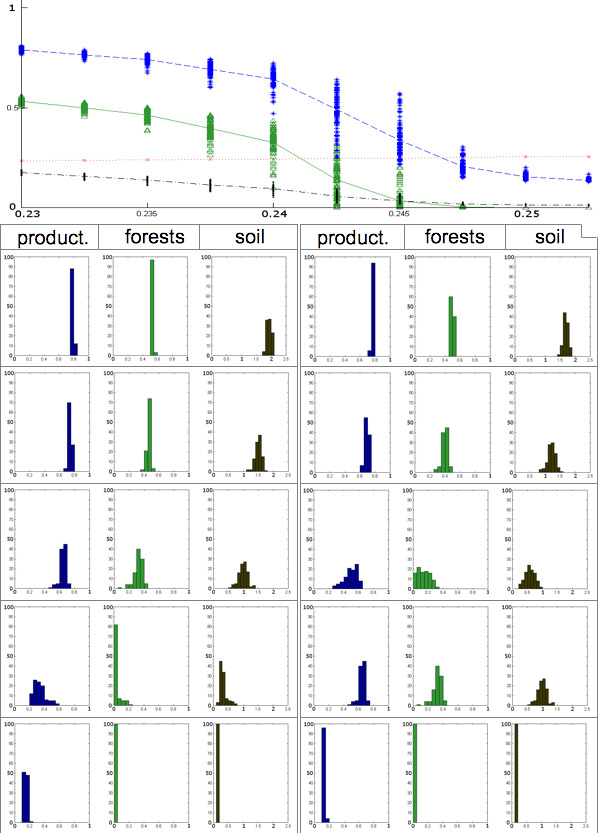}
  \put(-10,353.5){$\rightarrow$}\put(-10,346){$\gamma_{\rm agr}$}
  \put(-332,320){$0.23$}\put(-160,320){$0.2325$}
  \put(-332,253){$0.235$}\put(-160,253){$0.2375$}
  \put(-332,188){$0.24$}\put(-160,188){$0.2425$}
  \put(-332,121){$0.245$}\put(-160,121){$0.2475$}
  \put(-332,54){$0.25$}\put(-150,54){$0.2525$}
  \caption[Detailed analysis of the transition region]{Detailed analysis of the transition region, performed
  for strategy $\text{[S1]}_{\text{rand}}^{\text{(fix)}}$. This figure contains an overview plot, structured
  as the one in fig.~\ref{fig:strat1rf}, but with 100 runs per single value of
  $\gamma_{\rm agr}\in\{0.23,\,0.2325,\,\ldots,\,0.25,\,0.2525\}$.
  In addition, for these values we show histogram plots, analogous to those presented
  in fig.~\ref{fig:init_cond}. These are plots with 20 bins in $[0,\,1]$ or normalized
  production rate and for forest density, histogram plots with 25 bins in $[0,\,2.5]$
  for the average amount of soil.}
  \label{fig:histograms}
\end{figure}
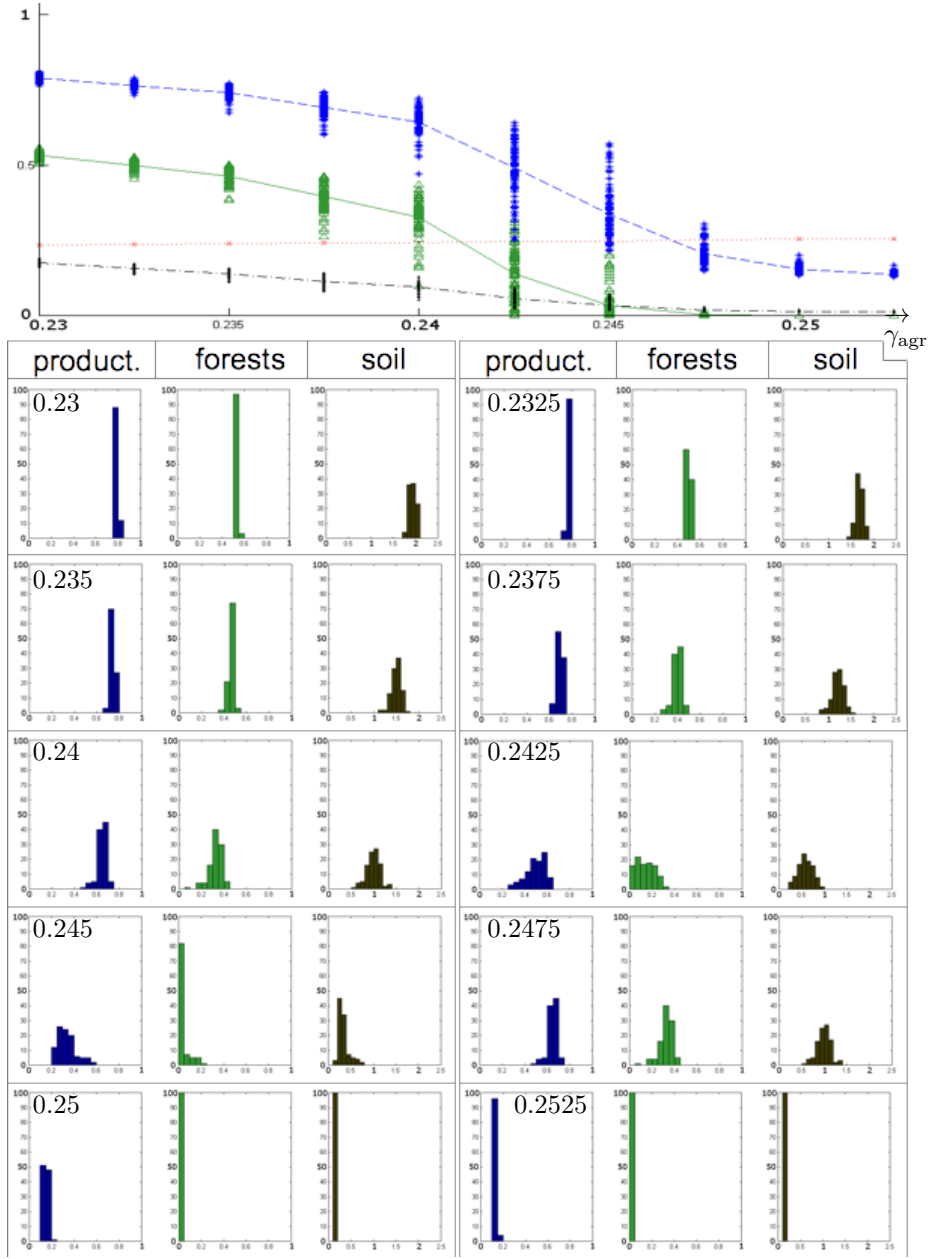

This region has been analyzed in more detail. The corresponding overview plot and some
histograms are displayed in figure~\ref{fig:histograms}. In particular for the target values
$\gamma_{\rm agr}= 0.2425$ and $\gamma_{\rm agr}= 0.245$ the spread of production
rates is large.

\medskip

Two effects contribute to this spread: Both the basic update rules and (in this case) the strategy
for the selection of new agricultural cells contain random elements. While these stochastic
aspect seems to be of little importance for regions of clear success (here $\gamma_{\rm agr}<0.23$)
and regions of clear failure (here $\gamma_{\rm agr}>0.25$), random events or choices may
strongly influence the system in the critical region.

In addition, it is likely (though presumably very difficult to prove) that -- beyond the influence
of stochastic events -- the update rules also contain aspects of chaotic behaviour. In particular the
neighbourhood-dependent update rule for forests described by~\eqref{eq:update_forest} may
lead to situations where survival of the forest critically depends not only on the density, but also
on the precise distribution of forest cells. In such cases, the initial configuration (which is also
generated using random numbers) can be of crucial importance for the outcome.

\begin{figure}
  \includegraphics[width=12cm]{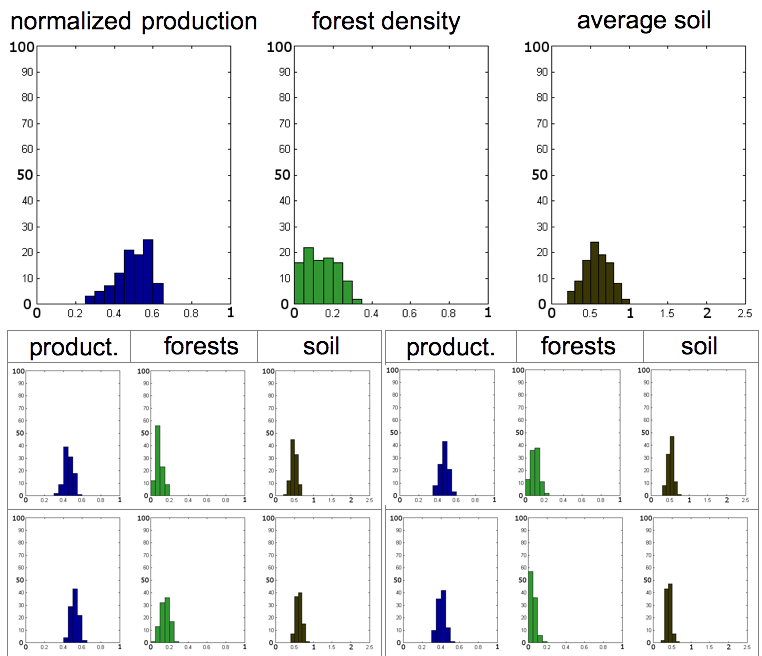}
  \put(-352,281){(a)}\put(-352,136){(b)}
  \caption[Influence of initial conditions]{Influence of initial conditions,
    performed for strategy $\text{[S1]}_{\text{rand}}^{\text{(fix)}}$.\\[3pt]
    (a) Histogram for hundred
    different simulation runs for $\gamma_{\rm agr}=0.2425$ (the same histogram as the one already
    shown in fig.~\ref{fig:histograms} for this value of $\gamma_{\rm agr}$, but enlarged).\\[3pt]
    (b) Histograms for four sets of simulation runs for $\gamma_{\rm agr}=0.2425$ with hundred
    runs per set, but this time for each distinct set only one (randomly chosen) initial configuration
    has been used. For each set, the spread of results can only be attributed to the dynamics of the
    system, not to sensitivity to initial conditions.\\[3pt]
    As in fig.~\ref{fig:histograms}, the histograms contain $20$ bins each, uniformly placed in the
    interval $[0,\,1]$ for normalized production rate and for forest density, $25$ bins uniformly
    placed in $[0,\,2.5]$ for the average amount of soil.}
  \label{fig:init_cond}
\end{figure}

In order to obtain a rough estimate of the importance of this effect, we have performed
multiple simulation runs for the same initial configuration. The results (histograms for 100
runs with $\gamma_{\rm agr}= 0.2425$ per initial configuration) for four distinct initial
conditions are shown in fig.~\ref{fig:init_cond}.

While the spread for runs with the same initial configuration is significantly smaller than for runs
with different configurations, it is still present. This is at least some evidence (though by far no proof)
that both random fluctuations and chaotic behaviour contribute to the spread of results in the
transition region.

\subsection{Robustness of the Model}
\label{ssec:sample_parchoice}

All simulations so far have been performed with one set of parameters, given in
tab.~\ref{tab:standardpar}. For the model to be relevant, the qualitative behaviour
should not depend heavily on the actual choice of parameters, as long as, for example,
the ordering $s^{(+)}_{(\text{pio})} < s^{(+)}_{(\text{for})} < s^{(-)}_{(\text{agr})}$
is respected.

In order to check this for at least a few cases, in addition to tab.~\ref{tab:standardpar}
we have defined two other sets of parameters\footnote{Since only the ratios of the five
parameters  $\tau_{\text{for}}$, $\tau_{\rm agri} = 1$, $s^{(+)}_{(\text{pio})}$,
$s^{(+)}_{(\text{for})}$ and $s^{(-)}_{(\text{agr})} = 0.1$, one of these parameters
can be chosen arbitrarily, which sets the absolute scale. We have chosen to keep
$\tau_{\rm agri}=1$, as in all other simulations, and vary the other parameters.},
given in tab.~\ref{tab:modpar} and created systematic plots similar to the one
shown in fig.~\ref{fig:strat1rf}.

\begin{table}
  \begin{center}
  \begin{tabular}{|llll|} \hline
  grid size \& timesteps & $N_1=32$, & $N_2=32$, & $N_t=256$, \\[3pt] \hline
  first modified set: &
  $\rho_{\text{for}}^{\text{(ini)}} = 0.25$, &
  $\rho_{\text{pio}}^{\text{(ini)}} = 0.25$, & \\[6pt]
  & $p_{\text{pio}} = 0.25$, &
  $\tau_{\text{for}} = 0.75$, &
  $\tau_{\rm agri} = 1$, \\[6pt]
  & $s^{(+)}_{(\text{pio})} = 0.02$, &
  $s^{(+)}_{(\text{for})} = 0.04$, &
  $s^{(-)}_{(\text{agr})} = 0.12$ \\[6pt] \hline
  second modified set: &
  $\rho_{\text{for}}^{\text{(ini)}} = 0.75$, &
  $\rho_{\text{pio}}^{\text{(ini)}} = 0.75$, & \\[6pt]
  & $p_{\text{pio}} = 0.75$, &
  $\tau_{\text{for}} = 0.25$, &
  $\tau_{\rm agri} = 1$, \\[6pt]
  & $s^{(+)}_{(\text{pio})} = 0.04$, &
  $s^{(+)}_{(\text{for})} = 0.06$, &
  $s^{(-)}_{(\text{agr})} = 0.08$ \\ \hline
\end{tabular}
\end{center}
\caption{Set of modified parameters, used for the simulations presented
in sec.~\ref{ssec:sample_parchoice} and fig.~\ref{fig:modified}.}
\label{tab:modpar}
\end{table}

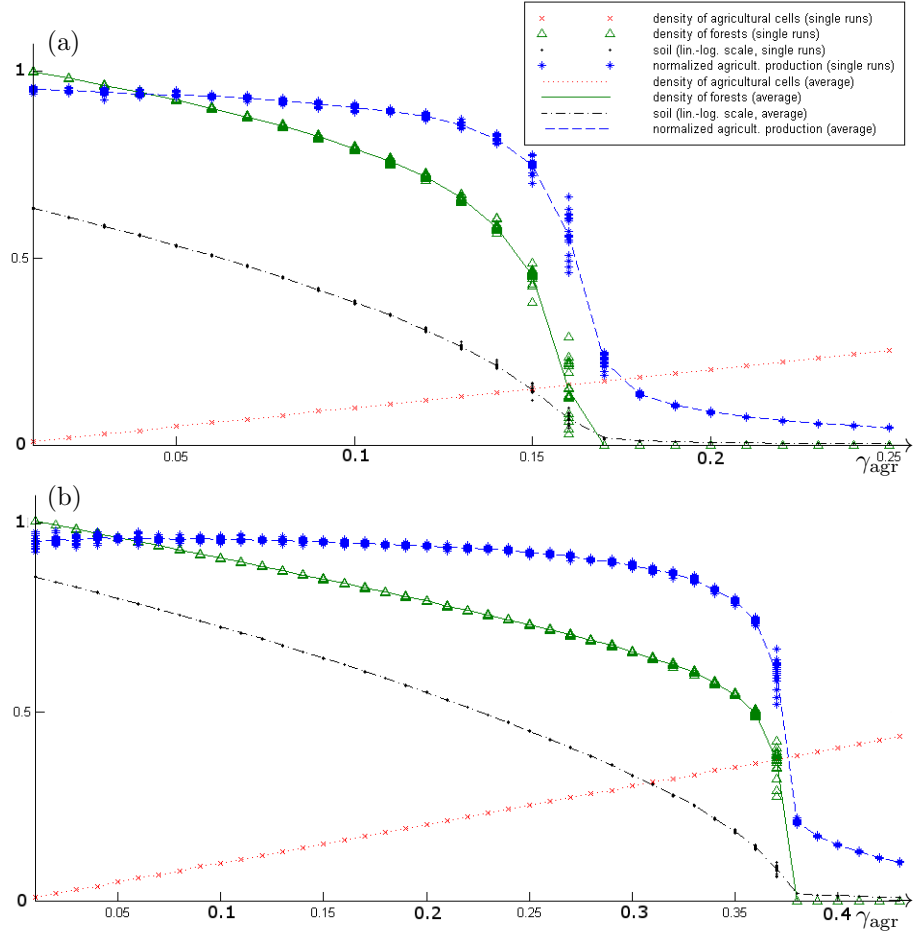
\begin{figure}
  \includegraphics[width=12cm]{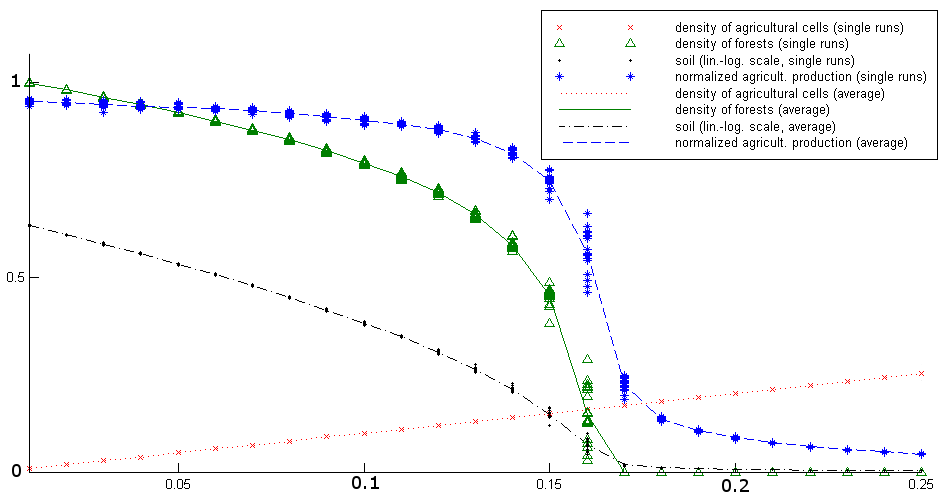}\put(-8,6){$\rightarrow$}\put(-20,0){$\gamma_{\rm agr}$}\put(-326,158){(a)}\newline
  \includegraphics[width=12cm]{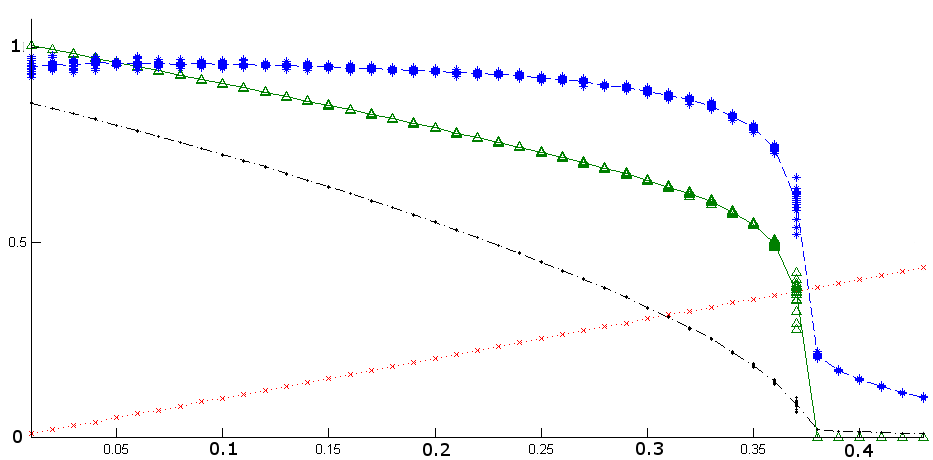}\put(-8,5.8){$\rightarrow$}\put(-20,0){$\gamma_{\rm agr}$}\put(-326,158){(b)}
  \caption[Overview plots for modified parameters.]{Overview plots for modified parameters:
     Simulation runs as in fig.~\ref{fig:strat1rf} (strategy $\text{[S1]}_{\text{rand}}^{\text{(fix)}}$,
     20 runs for each single value of $\gamma_{\rm agr}$), but with parameters taken from
     tab.~\ref{tab:modpar} and performed for a different range for  $\gamma_{\rm agr}$:\\
     (a) Results for first modified set of parameters, $\gamma_{\rm agr}\in 0.01\cdot\{1,\,2,\,\ldots,\,25\}$,\\
     (b) Results for second modified set of parameters, $\gamma_{\rm agr}\in 0.01\cdot\{1,\,2,\,\ldots,\,43\}$
     See fig.~\ref{fig:strat1rf} for a more detailed explanation of the quantities displayed in the plot.}
  \label{fig:modified}
\end{figure}

The plots for the two alternative sets of parameters is shown in fig.~\ref{fig:modified}.
While the transitions from success ($A\approx A^{(\rm goal)}$) to failure
($A\ll A^{(\rm goal)}$) set in at different values of $\gamma_{\rm agr}$,
the qualitative behaviour is essentially unaltered. This indicates that the model
is robust with respect to variations of parameters.

\section{Comparison of Strategies}
\label{sec:comparison}

All simulations in the sec.~\ref{sec:sample_analysis} have been performed with strategy
$\text{[S1]}_{\text{rand}}^{\text{(fix)}}$. We now turn to some other management
strategies outlined in sec.~\ref{sssec:resource_management}. The results for these simulations
are shown as a combination of a comprehensive overview plot (such as  the one shown in
fig.~\ref{fig:strat1rf}) and two or three single runs (similar to the one shown in
fig.~\ref{fig:run32-24-25}), combined into one single figure.

\subsection{Variable Number of Agricultural Fields}
\label{ssec:comparison_variable}

\begin{figure}
  \includegraphics[width=12cm]{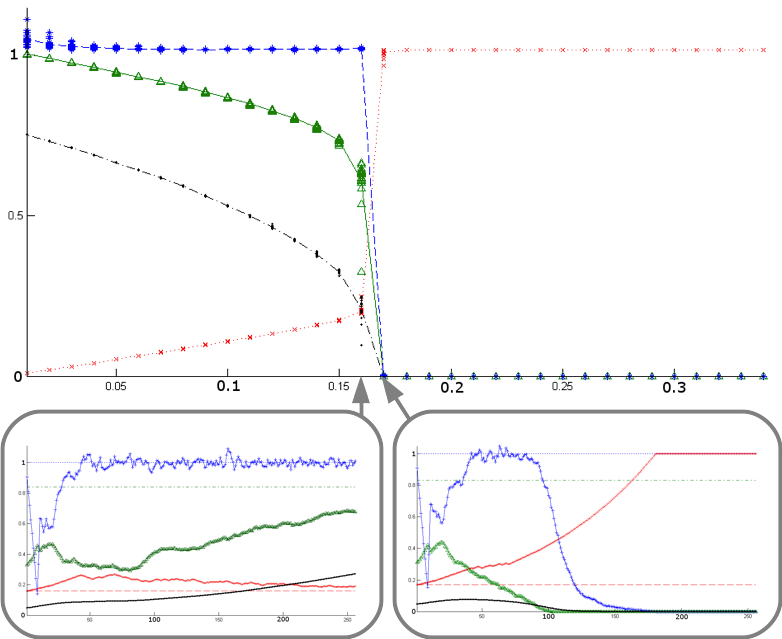}\put(-8,112.1){$\rightarrow$}\put(-20,106){$\gamma_{\rm agr}$}
  \put(-330,88){$\gamma_{\rm agr}=0.16$}\put(-160,88){$\gamma_{\rm agr}=0.17$}
  \caption[Overview plot for strategy $\text{[S1]}_{\text{rand}}^{\text{(var)}}$]
    {Overview plot for strategy $\text{[S1]}_{\text{rand}}^{\text{(var)}}$
     with thumbnails of typical single runs for $\gamma_{\rm agr}=0.16$ and
     $\gamma_{\rm agr}=0.17$ included.  The overview plot is structured the
     same way as the one in fig.~\ref{fig:strat1rf}, the single runs as the ones in
     fig.~\ref{fig:run32-24-25}.}
  \label{fig:strat1rv}
\end{figure}

When employing strategy $\text{[S1]}_{\text{rand}}^{\text{(var)}}$,
regarding the number of agricultural fields as variable, the transition between
success and failure becomes even more pronounced than for strategy
$\text{[S1]}_{\text{rand}}^{\text{(fix)}}$. For the standard choice of
parameters, the system exhibits dramatic breakdown between
$\gamma_{\rm agr}=0.16$ and $\gamma_{\rm agr}=0.17$,
see fig.~\ref{fig:strat1rv}.

This type of behaviour can easily be explained. For small values
of $\gamma_{\rm agr}$, the target production is easily reached and
the degree of freedom provided by a variable number of fields is not used.
For larger values of $\gamma_{\rm agr}$, when the production drops below
the target, additional fields are created.

In the present case, for $\gamma_{\rm agr}\le 0.16$, these additional fields
make up for the production losses, and full productivity can be provided.
For $\gamma_{\rm agr}\ge 0.17$, the increased number of fields reduces
the number of cells available for regeneration of soil, which leads to a further
increase of the number of fields -- a positive feedback loop which leads
to breakdown of the whole system and zero productivity.

In general, this type of behaviour can also be expected for other strategies
with a variable number of fields. The transition will be more pronounced
and failure will occur already at lower values of $\gamma_{\rm agr}$ than
for the otherwise equivalent strategy with a fixed number of fields.

\subsection{Optimized Choice of New Agricultural Fields}
\label{ssec:comparison_optimized}

In strategy $\text{[S1]}_{\text{opt}}^{\text{(fix)}}$, the number of fields
is kept fixed, but instead of beeing chosen randomly, new fields are chosen in
an optimized way. This means that those cells with maximum amount of soil
are converted to fields.

\begin{figure}
  \includegraphics[width=12cm]{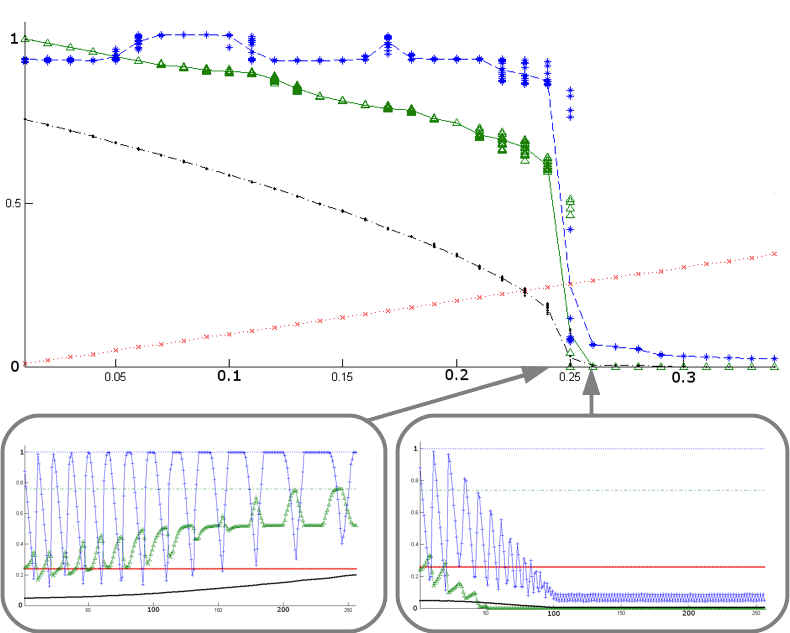}\put(-8,112.5){$\rightarrow$}\put(-20,106){$\gamma_{\rm agr}$}
  \put(-330,85){$\gamma_{\rm agr}=0.24$}\put(-160,85){$\gamma_{\rm agr}=0.26$}
  \caption[Overview plot for strategy $\text{[S1]}_{\text{opt}}^{\text{(fix)}}$]
    {Overview plot for strategy $\text{[S1]}_{\text{opt}}^{\text{(fix)}}$
     with thumbnails of typical single runs for $\gamma_{\rm agr}=0.24$ and
     $\gamma_{\rm agr}=0.26$ included.  The overview plot is structured the
     same way as the one in fig.~\ref{fig:strat1rf}, the single runs as the ones in
     fig.~\ref{fig:run32-24-25}.}
  \label{fig:strat1of}
\end{figure}

The results are shown in fig.~\ref{fig:strat1of}. For $\gamma_{\rm agr}<0.25$,
strategy works well, though there are some variations in productivity, as
discussed below. The transition at $\gamma_{\rm agr}=0.25$ is sharp, but
for this value one has a large spread of possible outcomes. While most
simulation runs for this target fail, some are reasonable successful with
a normalized productivy in the range of $80\%$ to $90\%$.

For this strategy, a close look at single runs is particularly rewarding.
Both in fig.~\ref{fig:strat1of}.b and~\ref{fig:strat1of}.c one has
significant oscillations of productivity. While the first few periods are
quite similar in both cases, soon the difference becomes clearly visible.
For fig.~\ref{fig:strat1of}.c, the height of the peaks drops and the
average amount of forest during a cycle decreases. For
fig.~\ref{fig:strat1of}.b, the duration of maximum productivity
during a single cycle increases (i.e. plateaus form).

However, steep drops in productivity occur also after considerable
simulation time and while these drops seem to become less severe
during the course of simulation, this happens only at a very slow pace.

These oscillations are, to a lesser extent, still present also for smaller values of
$\gamma_{\rm agr}$. This is visible in fig.~\ref{fig:strat1of}.a: Depending on
whether or not such a characteristic drop in productivity is included in the
averaging region (i.e. the last $64$ timesteps), the averaged production is either
very close to the target or about $5\%$ below.

\subsection{Optimized Choice with Variable Number of Fields}
\label{ssec:comparison_optimized_var}

\begin{figure}
  \includegraphics[width=12cm]{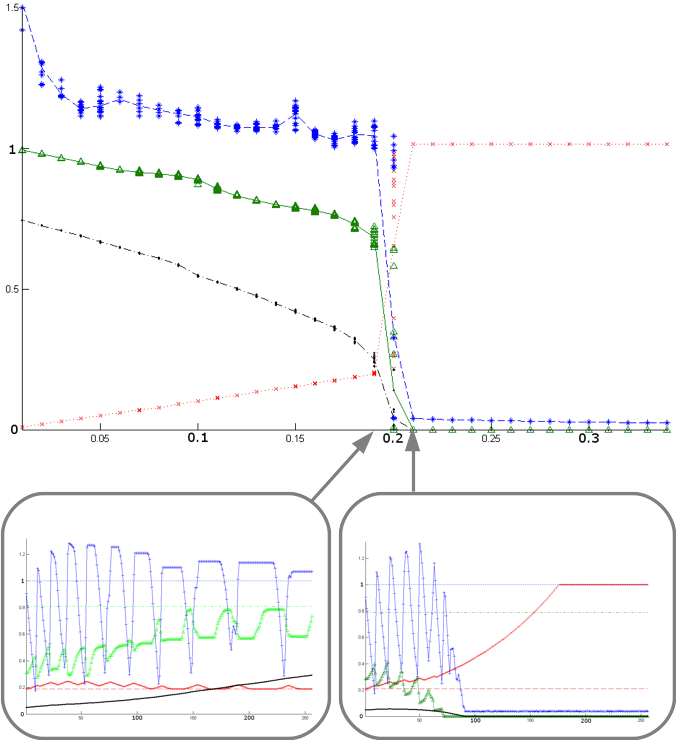}\put(-8,154){$\rightarrow$}\put(-20,148){$\gamma_{\rm agr}$}
  \put(-330,112){$\gamma_{\rm agr}=0.19$}\put(-160,112){$\gamma_{\rm agr}=0.21$}
  \caption[Overview plot for strategy $\text{[S1]}_{\text{opt}}^{\text{(var)}}$]
    {Overview plot for strategy $\text{[S1]}_{\text{opt}}^{\text{(var)}}$
     with thumbnails of typical single runs for $\gamma_{\rm agr}=0.19$ and
     $\gamma_{\rm agr}=0.21$ included.  The overview plot is structured the
     same way as the one in fig.~\ref{fig:strat1rf}, the single runs as the ones in
     fig.~\ref{fig:run32-24-25}.}
  \label{fig:strat1ov}
\end{figure}

Strategy $\text{[S1]}_{\text{opt}}^{\text{(var)}}$ combines the modifications
of secs.~\ref{ssec:comparison_variable} and~\ref{ssec:comparison_optimized},
i.e. new fields are chosen according to maximum soil and the number of fields
is variable. Results are shown in fig.~\ref{fig:strat1ov}, and as already stated
in sec.~\ref{ssec:comparison_variable}, a variable number of fields leads to
a sharp transition and an earlier onset of failure. This strategy permits ``over-shooting''
the production goal, and for small values of $\gamma_{\rm agr}$, the normalized
productivity tends to significantly exceed one.

As it had already been the case in sec.~\ref{ssec:comparison_optimized},
the single runs displayed as thumbnails in figs.~\ref{fig:strat1ov} exhibit
prominent cycles both in the case of success and of failure. In both
cases a typical cycle includes a phase of over-production and a steep drop
to $30\%$ productivity or less.

The difference between the two cases is not obvious during the first few cycles
when just focussing on productivity. Also the number of fields and the amount of
soil show strikingly similar behaviour. The best early indicator for success or failure
seems to be the net increase or decrease of forest during a single cycle.

\subsection{Abandoning Fields Below Threshold}
\label{ssec:comparison_abandon}

An alternative to keeping fields as long as possible is to abandon fields already if
the amount of soil drops below threshold. This is implemented in the different
variations of stategy $\text{[S2]}$. One can expect that this strategy can avoid
drops in productivity better than strategy $\text{[S1]}$, as long as sufficient
resources (i.e. cells with a sufficient amount of soil) are available.

\begin{figure}
  \includegraphics[width=12cm]{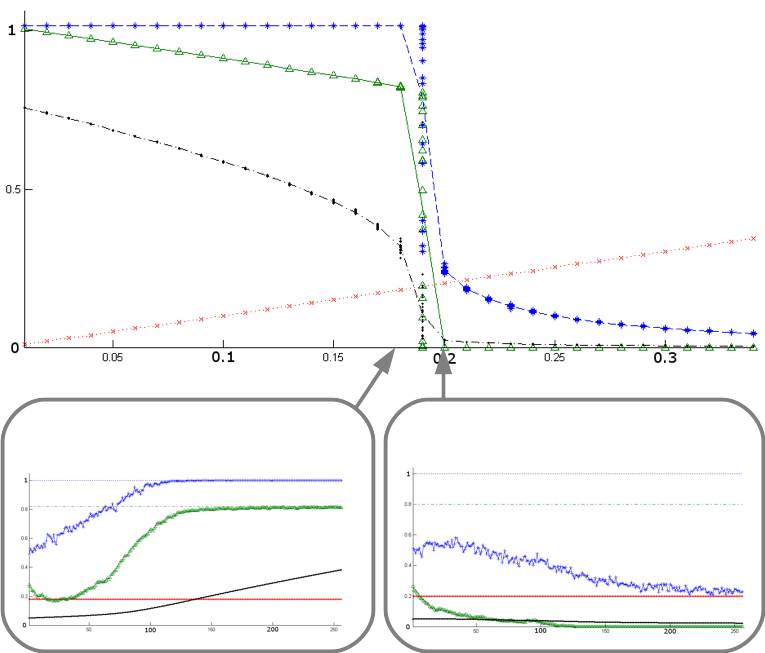}\put(-8,134){$\rightarrow$}\put(-20,128){$\gamma_{\rm agr}$}
  \put(-330,95){$\gamma_{\rm agr}=0.18$}\put(-160,95){$\gamma_{\rm agr}=0.20$}
  \caption[Overview plot for strategy $\text{[S2]}_{\text{rand}}^{\text{(fix)}}$]
    {Overview plot for strategy $\text{[S2]}_{\text{rand}}^{\text{(fix)}}$
     with thumbnails of typical single runs for $\gamma_{\rm agr}=0.18$ and
     $\gamma_{\rm agr}=0.20$ included.  The overview plot is structured the
     same way as the one in fig.~\ref{fig:strat1rf}, the single runs as the ones in
     fig.~\ref{fig:run32-24-25}.}
  \label{fig:strat2rf}
\end{figure}

As an example for this type of strategies, we have examined
strategy $\text{[S2]}_{\text{rand}}^{\text{(fix)}}$; the results are displayed
in fig.~\ref{fig:strat2rf}. As compared to the results for the otherwise
equivalent strategy $\text{[S1]}_{\text{rand}}^{\text{(fix)}}$,
displayed in fig.~\ref{fig:strat1rf}, on has a more stable production rate for
$\gamma_{\rm agr}\le 0.18$, but a sharp transition (with a wide spread of possible
outcomes for the same target productivity) at $\gamma_{\rm agr}= 0.19$.

As it has already been the case with strategies $\text{[S1]}_{\text{opt}}^{\text{(fix)}}$,
a seemingly ``smarter'' strategy (corresponding to more efficient exploitation) leads to
a sharper transition and an earlier breakdown of the system.

\subsection{Redistribution of Fields} 
\label{ssec:comparison_redistributionwithout}

\begin{figure}
  \includegraphics[width=12cm]{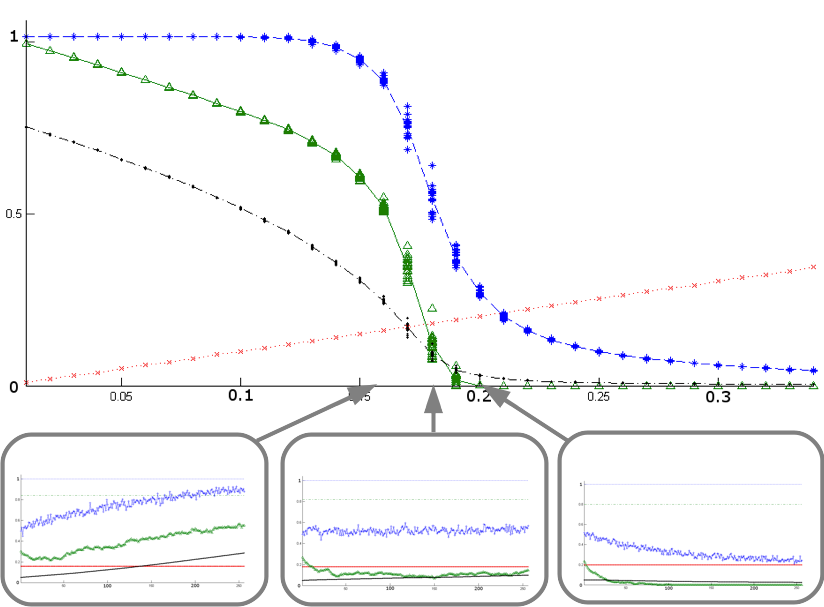}\put(-8,89.8){$\rightarrow$}\put(-20,86){$\gamma_{\rm agr}$}
  \put(-330,61){$\gamma_{\rm agr}=0.16$}
  \put(-216,61){$\gamma_{\rm agr}=0.18$}
  \put(-102,61){$\gamma_{\rm agr}=0.20$}\\
  \caption[Overview plot for strategy $\text{[S3]}_{\text{rand}}^{\text{(fix)}}$]
    {Overview plot for strategy $\text{[S3]}_{\text{rand}}^{\text{(fix)}}$
     with thumbnails of typical single runs for $\gamma_{\rm agr}=0.16$,
     $\gamma_{\rm agr}=0.18$ and $\gamma_{\rm agr}=0.20$ included.
     The overview plot is structured the same way as the one in fig.~\ref{fig:strat1rf},
     the single runs as the ones in fig.~\ref{fig:run32-24-25}.}
  \label{fig:strat3rf}
\end{figure}

Strategy $\text{[S3]}_{\text{rand}}^{\text{(fix)}}$ includes random redistribution
of fields at each timestep. The results are illustrated in fig.~\ref{fig:strat3rf}.
Compared to the otherwise equivalent strategies $\text{[S1]}_{\text{rand}}^{\text{(fix)}}$
and $\text{[S2]}_{\text{rand}}^{\text{(fix)}}$, one has a smoother transition
between high and low productivity.

The normalized production has dropped to about $50\%$ at $\gamma_{\rm agr}\approx 0.18$,
which is similar to the transition value found for $\text{[S2]}_{\text{rand}}^{\text{(fix)}}$
(see fig.~\ref{fig:strat2rf}) and significantly lower than the one found for
$\text{[S1]}_{\text{rand}}^{\text{(fix)}}$ (see fig.~\ref{fig:strat1rf}).

\subsection{Redistribution of Fields with Reforestation}
\label{ssec:comparison_redistributionwith}

\begin{figure}
  \includegraphics[width=12cm]{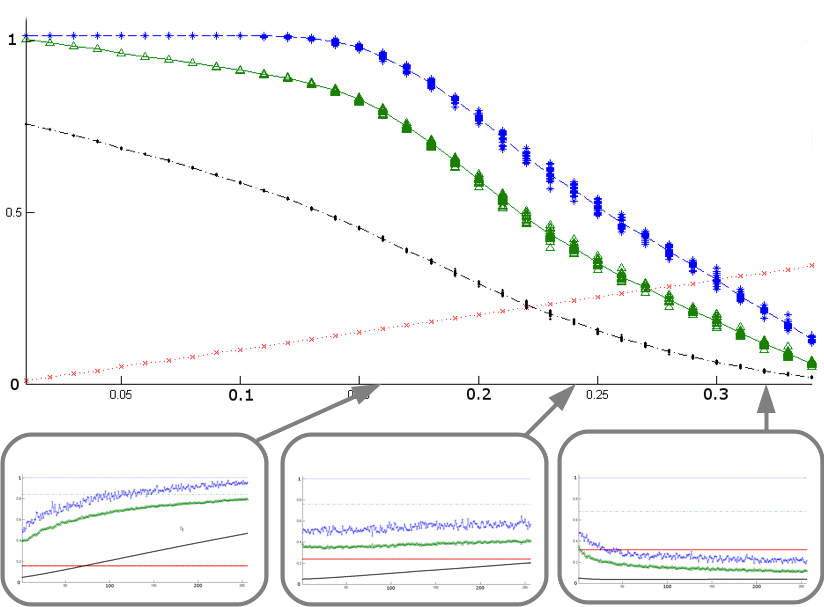}\put(-8,89.8){$\rightarrow$}\put(-20,86){$\gamma_{\rm agr}$}
  \put(-330,61){$\gamma_{\rm agr}=0.16$}
  \put(-216,61){$\gamma_{\rm agr}=0.24$}
  \put(-102,61){$\gamma_{\rm agr}=0.32$}\\
  \caption[Overview plot for strategy $\text{[S4]}_{\text{rand}}^{\text{(fix)}}$]
    {Overview plot for strategy $\text{[S4]}_{\text{rand}}^{\text{(fix)}}$
     with thumbnails of typical single runs for $\gamma_{\rm agr}=0.16$,
     $\gamma_{\rm agr}=0.24$ and $\gamma_{\rm agr}=0.32$ included.
     The overview plot is structured the same way as the one in fig.~\ref{fig:strat1rf},
     the single runs as the ones in fig.~\ref{fig:run32-24-25}.}
  \label{fig:strat4rf}
\end{figure}

In strategy $\text{[S4]}_{\text{rand}}^{\text{(fix)}}$, redistribution of fields
is supplemented by reforestation, if possible; results are shown in fig.~\ref{fig:strat4rf}.
The transition is even less dramatic than the one illustrated in fig.~\ref{fig:strat3rf},
and one still has reasonable productivity at high values of $\gamma_{\rm agr}$.
The predefined production goal, however, can only be fulfilled for $\gamma_{\rm agr}\le 0.15$,
and it is clear from the graph that total productivity is maximum for an intermediate
value of $\gamma_{\rm agr}$ and drops by further increasing the target production.

\section{Discussion and Interpretation}
\label{sec:dicuss_interpret}

The model described in sec.~\ref{sec:model} and examined in secs.~\ref{sec:sample_analysis}
and~\ref{sec:comparison}, while being quite simple, exhibits extremely interesting behaviour.
In particular, for some strategies, as the ones labeled as $\text{[S1]}$ and $\text{[S2]}$ in
sec.~\ref{sssec:resource_management}, one can have dramatic drops in productivity when just
slightly raising the productivity goal above a critical value. A strategy that works very well
with a specific target productivit can yield extremely poor results with a higher one.

A particularly interesting strategy is $\text{[S1]}_{\text{opt}}^{\text{(var)}}$, discussed in
sec.~\ref{ssec:comparison_optimized_var}. It is to some extent ``intelligent'' for choosing
new fields in an optimized way and ``flexible'' for expanding the number of fields, if necessary.
It is also ``stubborn'' since it does not give up fields even if productivity drops and
``short-sighted'' for not having a built-in sustainability strategy (in contrast to
strategy $\text{[S4]}$) and putting no upper limit on the number of fields.

The properties of being ``intelligent'', ``flexible'', ``stubborn'' and ``short-sighted'' at the same
time makes this strategy an excellent candidate model for typical human behaviour. Indeed, the
patterns displayed in figs.~\ref{fig:strat1ov}.b and~\ref{fig:strat1ov}.c qualitatively resemble
upturn and downturn cycles, most prominently known from economy. These cycles either leading
to stabilization on a high level or to breakdown of the system.

Other strategies, as the ones labeled as $\text{[S3]}$ and $\text{[S4]}$ in
sec.~\ref{sssec:resource_management}, are more stable in the sense that transitions are
not so dramatic (though the domain of acceptable productivity does not need to be larger).
Employing these strategies (re-distributing fields, possibly even with reforestation)
means larger effort. In a more detailed approach, this effort could be taken into account
by additional costs which reduce the net productivity.

\bigskip

While the model presented in sec.~\ref{sec:model} has been formulated in terms
of soil, forests and agricultural fields, it might serve as a quite general model for
use and overuse of renewable resources. It might even be possible, to re-interpret
it in other fields, as in psychological context. A simple model for burnout
could be obtained with the substitution:

\begin{center}
\begin{tabular}{|clc|}\hline
soil & $\to$ & personal resources \\
pioneer plants & $\to$ & recreational time \\
forest & $\to$ & social activities \\
fields & $\to$ & working hours \\
production goal & $\to$ & prescribed workload \\ \hline
\end{tabular}
\end{center}

An alternative re-interpretation could be done for exampe in economic context,
for example with the substitution:

\begin{center}
\begin{tabular}{|clc|}\hline
soil & $\to$ & existing infrastructure \\
pioneer plants & $\to$ & private initiatives \\
forest & $\to$ & community-based activities \\
fields & $\to$ & exploitation of infrastructure \\
production goal & $\to$ & expected profit \\ \hline
\end{tabular}
\end{center}

\section{Summary and Outlook}

We have presented an extended stochastic cellular automata model
for the management of renewable resources, formulated in terms of soil,
vegetation and agriculture (sec.~\ref{sec:model}), but general enough to
be interpreted in various other ways (sec.~\ref{sec:dicuss_interpret}).

On small grids the typical behaviour becomes clear already after a few
hundred simulation steps (sec.~\ref{ssec:longterm}), and the model
seems to be robust with respect to finite size effects (sec.~\ref{ssec:finite_size}).
Within a wide range, a change of parameters only leads to quantitative,
but not to qualitative changes (sec.~\ref{ssec:sample_parchoice}) of the results.

The behaviour of the model is strongly influenced by the management
strategy employed (sec.~\ref{sssec:resource_management}).
Simple strategies typically lead to sharp transitions between
fulfillment of goals and a breakdown of the system (sec.~\ref{sec:sample_analysis}
and secs.~\ref{ssec:comparison_variable} to~\ref{ssec:comparison_abandon}),
possibly with several cycles before one or the other stable state is reached.

From these results one can conclude that a strategy which works very well for a
specific target productivity may yield extremely poor results with respect to a higher
one. Close to the critical value, raising the productivity even slightly higher can lead
to desaster.

We also see that if more sophisticated strategies are chosen
(secs.~\ref{ssec:comparison_redistributionwithout}
and~\ref{ssec:comparison_redistributionwith}), less dramatic transitions occur.
We note, however, that applying such strategies requires a larger effort. This could
be taken into account by considering additional costs which reduce the net productivity.

\medskip

The model offers many possibilities for further studies and for extensions.
The strategies presented in sec.~\ref{sssec:resource_management} and investigated
in sec.~\ref{sec:comparison} are far from being exhaustive, and it might be worthwile
to examine more elaborate management strategies, in particular with a variable
but limited number of agricultural cells.

Possible extensions that could be implemented include nature protection areas and
elements of economy. For the latter one can take into account food price, but also
the energy sector, since agriculture can act both as a source and as a sink of energy
available to society.

\medskip

Finally, one could include this model in a ``small-world'' climate model as the the one
described in~\cite{LiSchapp:2011CarbonClimate}, in which plant growth and large-scale
forest fires have significant influence on climate dynamics, while carbon dioxide
concentration and temperature limit plant growth rates. Employing the present model
would allow to include the carbon storage capacity of soil and impose an additional
soil-quality constraint on plant growth rates.

\end{document}